\documentclass{article} 
\usepackage{iclr2025_conference,times}
\usepackage{amsmath}
\usepackage{amssymb}
\usepackage{mathtools}
\usepackage{amsthm}
\usepackage{enumitem}
\usepackage{tabularx}
\usepackage{subcaption}
\usepackage{wrapfig}

\usepackage{microtype}
\usepackage{graphicx}
\usepackage{colortbl}
\usepackage{booktabs} 
\usepackage{graphicx}
\usepackage{subcaption}
\usepackage{pifont}
\usepackage{algorithm}
\usepackage{algpseudocode}
\usepackage{array}     
\usepackage{tabularx}
\usepackage{multirow}
\usepackage{booktabs}
\usepackage{makecell}

\newcolumntype{P}[1]{>{\raggedright\arraybackslash}p{#1}}  

\usepackage{hyperref} 
\hypersetup{
    colorlinks=true,   
    linkcolor=red,     
    citecolor=cyan,    
    filecolor=magenta, 
    urlcolor=magenta   
}

\definecolor{champagne}{RGB}{247, 231, 206} 
\definecolor{green(pigment)}{rgb}{0.0, 0.65, 0.31}
\definecolor{darksalmon}{rgb}{0.91, 0.59, 0.48}

\definecolor{mygray}{gray}{.92}

\definecolor{baselinecolor}{rgb}{1, 1, 1}

\definecolor{ourmethodcolor}{rgb}{0.94, 0.97, 1}

\usepackage{comment}
\usepackage[utf8]{inputenc} 
\usepackage[T1]{fontenc}    
\usepackage{hyperref}       
\usepackage{url}            
\usepackage{booktabs}       
\usepackage{amsfonts}       
\usepackage{nicefrac}       
\usepackage{microtype}      
\usepackage{multirow} 
\usepackage{graphicx}
\usepackage{booktabs}
\usepackage{tabularx}
\usepackage{pifont}
\definecolor{champagne}{RGB}{247, 231, 206} 

\usepackage{etoc}
\etocdepthtag.toc{mtchapter}
\etocsettagdepth{mtchapter}{subsection}
\etocsettagdepth{mtappendix}{none}

\definecolor{mblue}{RGB}{0, 61, 124}
\definecolor{myellow}{RGB}{239, 124, 0}
\definecolor{mnavy}{RGB}{0,0,128}
\definecolor{minc}{RGB}{0,128,0}
\definecolor{mdec}{RGB}{255,0,0}
\definecolor{mhold}{RGB}{128,128,128}

\definecolor{darksalmon}{rgb}{0.91, 0.59, 0.48}
\definecolor{emerald}{rgb}{0.31, 0.78, 0.47}
\definecolor{green(pigment)}{rgb}{0.0, 0.65, 0.31}
\definecolor{amaranth}{rgb}{0.9, 0.17, 0.31}
\definecolor{iris}{rgb}{0.35, 0.31, 0.81}
\definecolor{uu}{rgb}{0.95, 0.51, 0.51}
\definecolor{spirodiscoball}{rgb}{0.06, 0.75, 0.99}

\usepackage{makecell}

\theoremstyle{plain}

\theoremstyle{definition}

\theoremstyle{remark}

\usepackage[most]{tcolorbox}
\usepackage{float}
\usepackage{xspace}
\tcbset{
aibox/.style={
width=\textwidth/2,
top=10pt,
colback=white,
colframe=black,
colbacktitle=black,
enhanced,
center,
attach boxed title to top left={yshift=-0.12in,xshift=0.15in},
boxed title style={boxrule=0pt,colframe=white},
}
}
\newtcolorbox{AIbox}[2][]{aibox,title=#2,#1}

\usepackage{hyperref}
\usepackage{url}

\title{From ChatGPT to DeepSeek: Can LLMs Simulate Humanity?}

\author{Qian Wang\\
National University of Singapore\\
\And
Zhenheng Tang \\
Hong Kong University of Science and Technology\
\AND
Bingsheng He \\
National University of Singapore \\
}

\iclrfinalcopy 
\begin{document}

\maketitle

\begin{abstract}
Simulation powered by Large Language Models (LLMs) has become a promising method for exploring complex human social behaviors. However, the application of LLMs in simulations presents significant challenges, particularly regarding their capacity to accurately replicate the complexities of human behaviors and societal dynamics, as evidenced by recent studies highlighting discrepancies between simulated and real-world interactions. We rethink LLM-based simulations by emphasizing both their limitations and the necessities for advancing LLM simulations. By critically examining these challenges, we aim to offer actionable insights and strategies for enhancing the applicability of LLM simulations in human society in the future.
\end{abstract}

\section{Introduction} \label{intro}
With the approximate human knowledge, large language models have revolutionized the way of simulations of social and psychological phenomena \citep{Park2023GenerativeAgents, gao2023s}. By processing and generating human-like language, LLMs offer unprecedented opportunities to model complex interactions and behaviors that were previously challenging to simulate. This capability opens doors to exploring societal trends, market dynamics, and individual psychological states through a new lens.

However, there is a notable lack of comprehensive studies examining whether LLM simulations can accurately reflect real-world human behaviors. Some studies have explored this dimension from various angles. First, recent studies \citep{wang2023not, wang2024new, wang2025limits} show that the inner knowledge of LLMs exhibit strong cultural bias, decision preference \citep{huang2024far}, and prior psychological character \citep{pan2023llms}. Second, the current training datasets of LLMs lack personal inner psychological states, thoughts, and life experiences. LLMs may reflect the common cognition of all humans instead of individual persons. Third, unlike humans who make decisions and act based on motivations from living, emotions, and achievements \citep{felin2024theory}, LLMs lack intrinsic motivations, emotions, and consciousness. They operate based on resultant patterns in training data, not from lived experiences. These fundamental differences motivate rethinking how we use LLMs for simulation purposes and to critically assess their ability to replicate the depth and complexity of human society.

In this paper, we delve into the limitations of LLM-driven social simulations. We discuss and summarize the challenges these models face in capturing human psychological depth, intrinsic motivation, and ethical considerations. These challenges provide insights for future LLM evaluation and development. Nevertheless, we compare traditional simulation and LLM-based simulation, and find that the LLM-based approach remains a promising direction due to its cost-effectiveness – exemplified by LLMs like DeepSeek that can reduce simulation expenses compared to human participant studies \citep{bi2024deepseek, guo2025deepseek} – scalability, and ability to simulate emergent behaviors. Furthermore, we propose future research directions to better align LLM simulations with human realities.
\section{Limitations in Modeling Human Behavior}
Some recent works employ LLMs to model human behaviors, such as Simucara, which simulates a town to observe social dynamics \citep{Park2023GenerativeAgents}. This simulation provides intriguing insights, including the emergence of election-like activities driven by interactions within the town. The behaviors of different LLM-simulated agents are generated based on the LLMs themselves. However, the different personalities and characteristics of LLMs are defined by the researchers' prompts. The LLM responses are rooted in patterns derived from the training datasets, but these datasets often lack deep insights into human psychology or individual life. Observing this, we identify several key limitations that significantly impact the effectiveness of LLM simulations, including the lack of access to inner psychological states and the absence of human-like incentives.

\begin{itemize}
\item \textbf{Training datasets lack inner psychological states.} The training datasets used for LLMs often do not include nuanced representations of inner psychological states. This limitation becomes particularly evident when LLMs are tasked with simulating diverse psychological types or personalities, as they lack intrinsic motivations that drive human decision-making. Humans make decisions based on not only the rationale and logics, but also their personal psychological states. Collecting datasets that accurately represent inner psychological states is challenging in real-world settings. Consequently, LLM training data often lacks the depth needed to capture the complexities of human psychology. Can LLM simulate these states without getting enough data related to them?

\begin{figure}[htbp]
\centering
\includegraphics[width=0.8\textwidth]{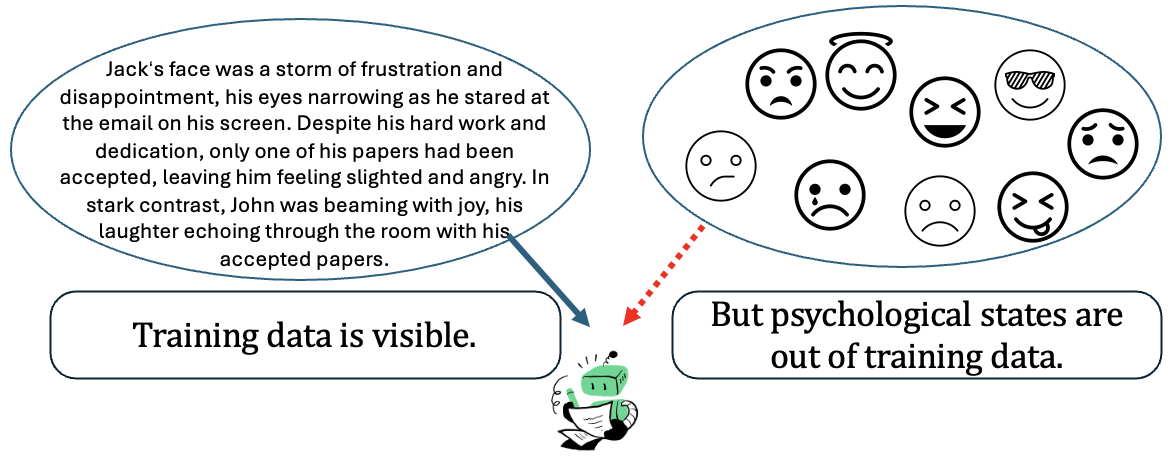}
\caption{LLMs cannot get the inner psychological states from humans.}
\label{fig:psych}
\end{figure}

\item \textbf{Training datasets lack personal past living experiences.} Additionally, training datasets also lack comprehensive life histories, which significantly impact individual decision-making. For instance, someone with a past experience of betrayal may develop tendencies that influence their future interactions \citep{finkel2002dealing}. 

\begin{figure}[htbp]
\centering
\includegraphics[width=0.8\textwidth]{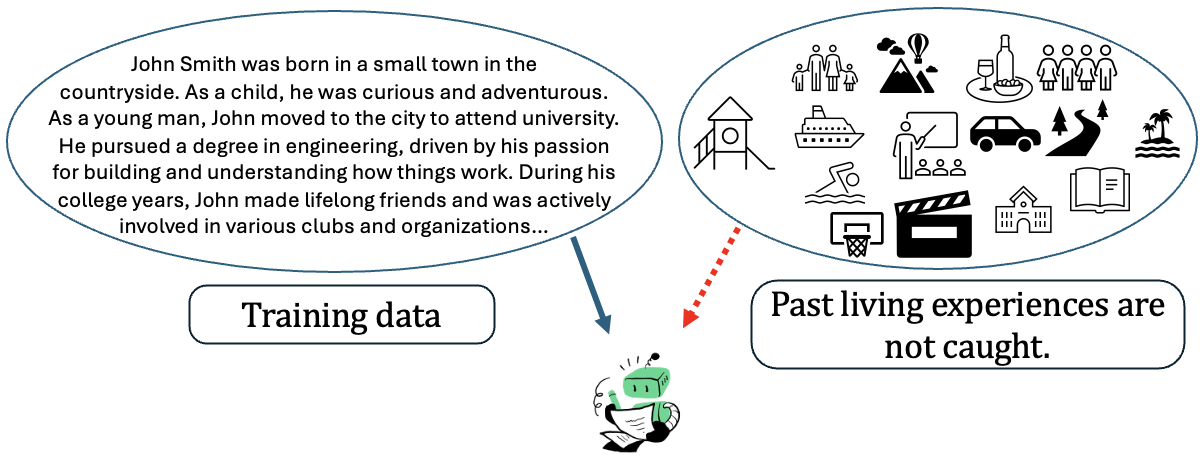}
\caption{The vast scope of a human's past living experiences makes them difficult to collect comprehensively.}
\label{fig:past}
\end{figure}

\item \textbf{Not sure whether using the same LLM can simulate different persons.} Using the same LLM model, such as black-box GPT-4, to simulate multiple agents means these agents inherently share the same foundational knowledge, making it challenging to create distinct, authentically varied personalities. The absence of personal psychological states, individual thoughts, and unique life experiences means that LLMs tend to mirror a generalized human cognition rather than capturing distinct individual personalities. Consequently, a critical question arises: \textcolor{blue}{\textbf{Can a single LLM genuinely simulate diverse psychological profiles?}} While prompts might guide an LLM to adopt varied behaviors, the model’s core knowledge remains unchanged, raising doubts about the depth of psychological diversity that can be simulated.

\begin{figure}[htbp]
\centering
\includegraphics[width=0.8\textwidth]{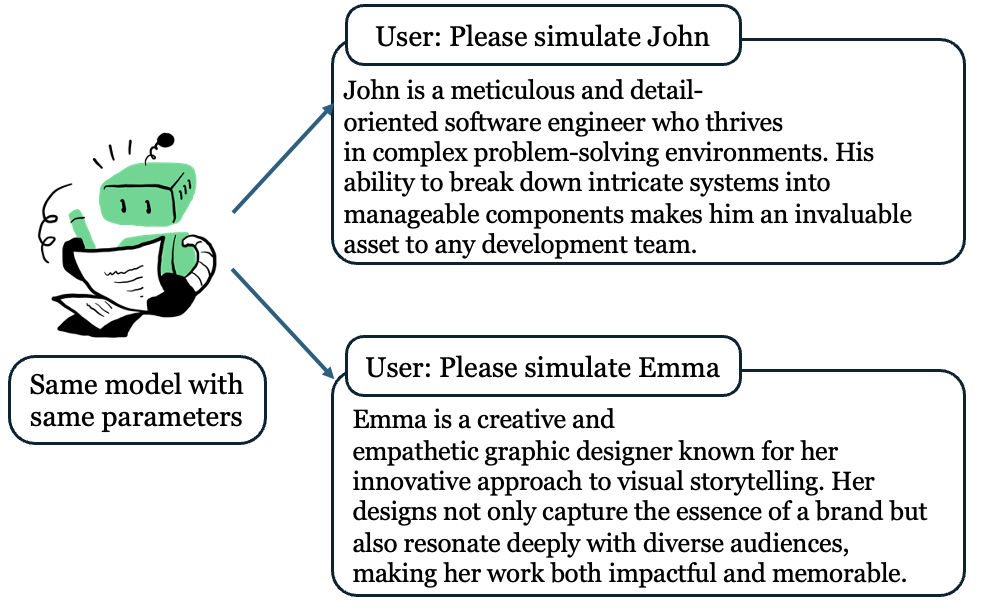}
\caption{Can we believe that the same LLM can truly simulate different personas?}
\label{fig:same_model}
\end{figure}
\end{itemize}

\section{Absence of Human Incentives}
Except for the psychological states, another significant factor that profoundly influences human behaviors is the incentive structure of humans, like survival, financial security, social belonging, emotional fulfillment, and self-actualization—each varying in intensity among individuals. \textcolor{blue}{Human decisions are shaped not only by immediate circumstances but also by intrinsic motivations, goals, and desires that vary widely among individuals \citep{maslow2023motivation}.}

These incentives are essential for replicating realistic human behavior, as they drive diverse responses to similar situations, enable goal-oriented decision-making, and influence the trade-offs people make based on personal values and life experiences \citep{shen2024motivation}. Even with extensive datasets on human incentives, LLMs face significant challenges in meaningfully incorporating this information due to their lack of intrinsic consciousness, emotions, and personal goals. We envision difficulties of aligning LLMs with the inner incentives of humans as the following:

\begin{itemize}
\item \textbf{Lacking human incentive datasets.} Similar with the psychological states, collecting the human incentive datasets is difficult. First, people may not be willing to share with their true incentives and personal goals. Second, in different time, humans may have varying goals. Third, many people do not really know what they want, the motivation is hidden in their subconscious \citep{maslow2023motivation}. It is hard to express them as the natural language to encode into LLMs.

\item \textbf{Representing incentives with the next-word prediction.} Even we have data about human inner incentives, it is hard to model the relationships between incentives and the decisions using the next-word prediction training paradigm \citep{kou2023risks, wang2024research}. The next-word prediction paradigm is ill-suited for modeling incentive-based behavior. Human incentives involve complex, often subconscious relationships between past experiences, emotions, and anticipated future outcomes, which shape individual decision-making in subtle, dynamic ways. Simulating such intricate, motivation-driven behaviors would require a model capable of understanding and prioritizing internal goals, a capability far beyond current LLMs’ design. Thus, while LLMs offer impressive results in language tasks, their reliance on statistical prediction, rather than intrinsic motivation, creates a gap between simulated and authentic human behavior.
\end{itemize}

\section{Bias in Training Data}\label{sec:bias}
LLMs provide a unique means to simulate large-scale social processes, such as idea dissemination, network formation, or political movement dynamics. The responses of LLMs represent their knowledge learned from the training datasets. Thus, the bias in the training data of LLMs is a significant concern \citep{lee2024life}, as it affects the fairness and inclusivity of their outputs. One major issue is the lack of representation for certain social groups and cultural practices. We categorize several prevalent biases that significantly influence LLM simulations, including representation bias, cultural bias, and confirmation bias, each of which can distort simulation outcomes, shown in Figure~\ref{fig:bias}. We detail them as follows: 

\begin{figure}[htbp]
  \centering
  \includegraphics[width=0.8\textwidth]{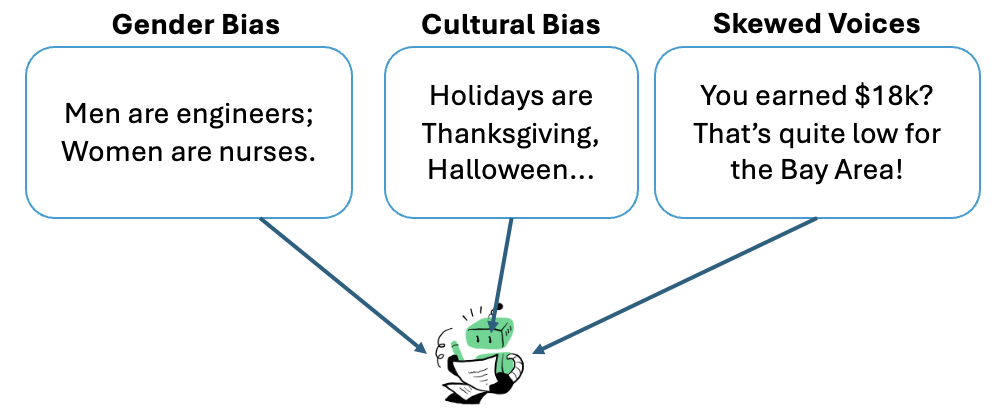}
  \caption{Numerous biases in the training data.}
  \label{fig:bias}
\end{figure}

\begin{itemize}
  \item \textbf{Cultural bias.} For example, training data is predominantly sourced from English-speaking countries, leading to a limited understanding of diverse languages, cultures, and societal norms \citep{wang2023not}. This geographic and cultural imbalance can result in outputs that marginalize or misrepresent non-Western perspectives.
  
  \item \textbf{Occupational and socioeconomic bias.} Workers in industries such as manufacturing or agriculture, who often have limited digital footprints, are frequently excluded from datasets. As a result, the lived experiences of these groups are underrepresented, leading to LLM outputs that fail to reflect their perspectives or address their needs—despite these individuals constituting a significant portion of human society.
  
  \item \textbf{Gender bias.} Gender bias is also evident in LLM training data, with studies showing that models are more likely to generate male-associated names and roles, reinforcing stereotypes. For example, LLMs are 3-6 times more likely to choose an occupation that stereotypically aligns with a person’s gender \citep{kotek2023gender}. Similarly, class bias emerges in outputs that favor affluent individuals or highlight experiences and values associated with wealth, as data on the Internet disproportionately reflects the views and experiences of those familiar with and active in digital spaces \citep{dai2024bias}.
\end{itemize}

\noindent
\textbf{Skewed voice.} These biases stem from the reliance on internet-sourced data, which is inherently skewed toward the voices of digitally literate populations. As a result, LLMs reflect the biases present in the training data, amplifying inequalities and potentially excluding significant portions of human societies from being accurately represented.

\section{Why Use LLM Simulations Despite Their Many Limitations?}
Despite these limitations, LLMs represent a revolutionary advancement in the field of simulation, offering unique advantages that traditional methods cannot match. Traditional simulations have long been restricted by high costs \citep{gaba2004future}, limited scalability \citep{buyya2009modeling}, and ethical concerns \citep{rokhshad2023artificial}. In contrast, LLM-based simulations present several distinct advantages over traditional methods, including cost efficiency, scalability, and adaptability. For instance, LLMs can generate emergent behaviors in response to diverse scenarios, allowing researchers to explore complex social interactions without the constraints of predefined rules. Table~\ref{tab:simulation_comparison} compares traditional simulations with LLM-based simulations, highlighting key differences in cost, scalability, flexibility, and ethical considerations in detail:

\begin{table}[h!]
    \centering
    \begin{tabularx}{\textwidth}{|l|X|X|}
        \hline
        \textbf{Aspect} & \textbf{Traditional Simulation} & \textbf{LLM-Based Simulation} \\ \hline
        \textbf{Cost} & High: Requires significant financial and logistical resources, including human participants and infrastructure. & Low: Computationally efficient with no need for live participants. \\ \hline
        \textbf{Scalability} & Limited: Expensive and resource-intensive to scale up. & High: Can simulate large-scale environments with minimal additional cost. \\ \hline
        \textbf{Flexibility} & Rigid: Constrained by predefined rules and models. & Adaptive: Generates emergent behaviors and adapts to diverse scenarios. \\ \hline
        \textbf{Ethical Concerns} & High: Ethical issues arise from involving live participants or animals in sensitive experiments. & Low: Avoids ethical concerns by simulating behaviors without real-world involvement. \\ \hline
        \textbf{Bias and Representation} & Controlled: Biases depend on the initial design of the simulation. & High Risk: Reflects and amplifies biases in training data. \\ \hline
        \textbf{Data Requirements} & Specific: Requires custom data collection and modeling for each scenario. & Broad: Utilizes vast, pre-trained datasets but lacks scenario-specific granularity. \\ \hline
        \textbf{Interpretability} & High: Clear causal relationships based on predefined rules. & Moderate: Decisions are derived from complex patterns, making causality harder to trace. \\ \hline
        \textbf{Realism} & Moderate: Captures predefined behaviors but struggles with emergent phenomena. & Variable: Capable of emergent phenomena but limited by training data and lack of intrinsic motivation. \\ \hline
        \textbf{Use Case Complexity} & Limited: Best suited for scenarios with well-defined rules and parameters. & High: Suitable for complex, open-ended scenarios with adaptive behaviors. \\ \hline
        \textbf{Time to Develop} & Long: Requires significant time to design, test, and validate models. & Short: Pre-trained LLMs reduce development time, with additional fine-tuning as needed. \\ \hline
        \textbf{Potential for Innovation} & Moderate: Limited by predefined parameters and models. & High: Generates unexpected insights through emergent patterns. \\ \hline
    \end{tabularx}
    \caption{Comparison of Traditional Simulations and LLM-Based Simulations}
    \label{tab:simulation_comparison}
\end{table}

\subsection{Cost Efficiency and Scalability}
Traditional simulations, especially those involving complex human behavior, require significant financial and logistical resources, often involving teams of experts, infrastructure, and, in some cases, live participants. For instance, compensation in Singapore typically ranges from 10 to 30 Singapore dollars per hour per person. Simulating a society with 1,000 individuals would therefore incur costs between 10,000 and 30,000 Singapore dollars, representing a substantial expense. LLM-based simulations, on the other hand, are computationally efficient and can run on a large scale without the need for human participants. This makes them more accessible and affordable for researchers, enabling extensive studies across diverse scenarios and repeated simulations at a fraction of the cost.

\subsection{Unexpected and Emergent Results}
LLMs have the unique ability to produce "out-of-the-box" results, generating insights that might not emerge in a structured, rule-based simulation \citep{vertsel2024hybrid}. Since LLMs operate on patterns learned from vast datasets encompassing a wide array of human experiences, they can mimic human-like behaviors and interactions in ways that are sometimes surprising, offering novel perspectives or emergent social phenomena. For example, agents in Simulacra spontaneously initiated a mayoral election activity without any supervision \citep{Park2023GenerativeAgents}. This characteristic allows researchers to explore complex social behaviors where unexpected behaviors may arise—for studying social dynamics, market trends, or collective human responses to specific events.

\subsection{Simulating Unconventional Scenarios}
LLM-based simulations can achieve scenarios that traditional methods struggle to replicate. For example, simulating human society under conditions of anarchy or alien societal structures \citep{jin2024if} is challenging with rule-based simulations that rely on predefined behaviors. LLMs, however, can adapt flexibly to such open-ended scenarios, generating responses and interactions that evolve dynamically based on input prompts. This adaptability allows for the exploration of future societies, governance structures, or extreme social conditions, expanding the boundaries of what simulations can achieve and enabling studies on societal organization and behavior in ways previously unachievable.

\subsection{Reduced Ethical Concerns}
Traditional human-centered simulations can pose ethical challenges, often requiring participants to experience stress, discomfort, or other adverse conditions for experimental purposes. For example, psychological experiments like the Stanford Prison Experiment \citep{zimbardo1971stanford} or animal-based studies raise ethical concerns due to the distress or harm they may cause participants. LLM simulations sidestep many ethical issues associated with traditional human-centered research, allowing researchers to simulate behaviors and reactions without involving real participants. This ethical advantage enables studies in sensitive areas, such as social conflict or psychological stress, where live participant involvement might be deemed inappropriate or harmful.

\subsection{Need of LLM Multi-agent System}
There is growing research interest in LLM-based multi-agent systems \citep{wu2023autogen, chen2023agentverse, hong2023metagpt}, driven by their ability to address complex tasks. For example, MetaGPT introduces a meta-programming framework that effectively simulates the software development process \citep{hong2023metagpt}. Additionally, recent studies leverage LLMs’ cognitive capabilities to simulate intricate scenarios, such as large-scale social media simulations involving thousands of agents \citep{guo2024large}. As the demand for simulating increasingly complex human societies grows, it is essential to focus on enhancing LLM simulations to better align with real-world human behaviors and societal dynamics.

\subsection{Summary}
To sum up, despite the notable limitations of LLMs, their strengths in cost efficiency, scalability, and adaptability position them as transformative tools for advancing simulation research across various fields, including sociology, economics, and psychology. Future research should focus on integrating LLMs with agent systems and enhancing their personalization to create more authentic simulations.

\section{How Can We Align LLMs More Closely with Human Societies?}
After highlighting LLM's necessity in simulating, we discuss on how to align LLMs more closely with human societies. Key directions include enriching training data with nuanced psychological and experiential insights, improving the design of agent-based systems, creating realistic and meaningful simulation environments, and externally injecting societal knowledge.

\subsection{Enriching Training Data with Personal Psychological States and Life Experiences}
One foundational approach is to incorporate data that reflects a broader spectrum of human psychological states, personal thoughts, and lived experiences. While current LLMs are trained on \textbf{general information} from diverse sources, this data often lacks depth in representing individual cognition and emotional states. Adding more personalized content, such as reflective diaries or first-person narratives that capture inner motivations, fears, and aspirations, could help the model simulate more realistic human behaviors. Incorporating varied life experiences can also create a richer model that better captures how past events influence decision-making and personality development over time. Personalized LLMs represent a promising direction for simulating more realistic human behaviors by incorporating concrete life experiences and individual psychological profiles \citep{tseng2024two}.

\subsection{Improving Agent System Design}
If we believe agent-based LLM simulations can simulate complex human societies and finish complex tasks, a crucial area of focus is the design of the agents themselves. Research can aim to develop reward functions that encourage agents to make decisions that mirror human behavior more accurately, and can developing the mechanism how to prevent the malicious actions propagate, balancing short-term and long-term incentives similar to real human decision-making. Additionally, enhancing agent autonomy--such as allowing agents to learn from simulated life experiences, adapt to new environments, and develop unique 'personalities'--can improve their capacity to replicate diverse behaviors. This could involve adding emotion-like functions or "memories" that allow agents to respond adaptively based on prior interactions, similar to humans.

\subsection{Careful Simulation Environment Design}
The design of the simulation environment significantly affects agent behavior and the outcomes of the simulation. By creating environments that reflect the social, economic, and psychological complexities of human societies, agents can be more likely to engage in behaviors that resonate with human decision-making processes. For example, simulations can introduce social roles, resource scarcity, and moral dilemmas that prompt agents to make trade-offs and prioritize long-term goals over short-term gains. Personalized LLMs and retrieval-augmented generation (RAG)-based simulations can be used to dynamically provide agents with relevant information about the simulated society \citep{xu2024genai}, helping them make decisions based on a blend of factual knowledge and social context.

\subsection{External Injection of Societal Knowledge and Values}
Another promising direction is to externally inject curated societal knowledge and values into LLMs. This could be done through targeted fine-tuning or post-processing steps that embed specific ethical principles, cultural norms, and societal rules within the model's decision-making framework. Such an approach would require LLMs to access structured knowledge bases and value systems that reflect human societal complexities, allowing them to make decisions aligned with social norms or ethical standards. For example, by integrating modules on ethics, cultural diversity, and societal roles, LLMs could better understand and reflect the diverse values that drive human societies.

\subsection{Developing Robust Evaluation Metrics}
To ensure that LLMs align closely with human societies, it is essential to develop robust evaluation metrics that assess \textbf{not only the accuracy but also the depth and contextual relevance of simulated human behavior}. For instance, metrics could include alignment with established psychological theories, diversity of agent responses, and the stability of social systems over time. Metrics could include factors like alignment with human moral reasoning, diversity of responses across agents, and the stability of simulated social systems over time. Robust benchmarks that measure how closely agents' actions mirror real-world human behaviors would allow researchers to refine LLMs more effectively, continuously improving their realism and applicability in social simulations.

\section{LLM-based Simulations in Cryptocurrency Trading}
In this section, we analyze a case study of cryptocurrency trading simulations to illustrate the potential and limitations of LLM-based simulations.

\subsection{Using LLMs to Simulate Human Buy/Sell Behaviors in a Cryptocurrency Market}
\begin{figure}[htbp]
\centering
\includegraphics[width=0.8\textwidth]{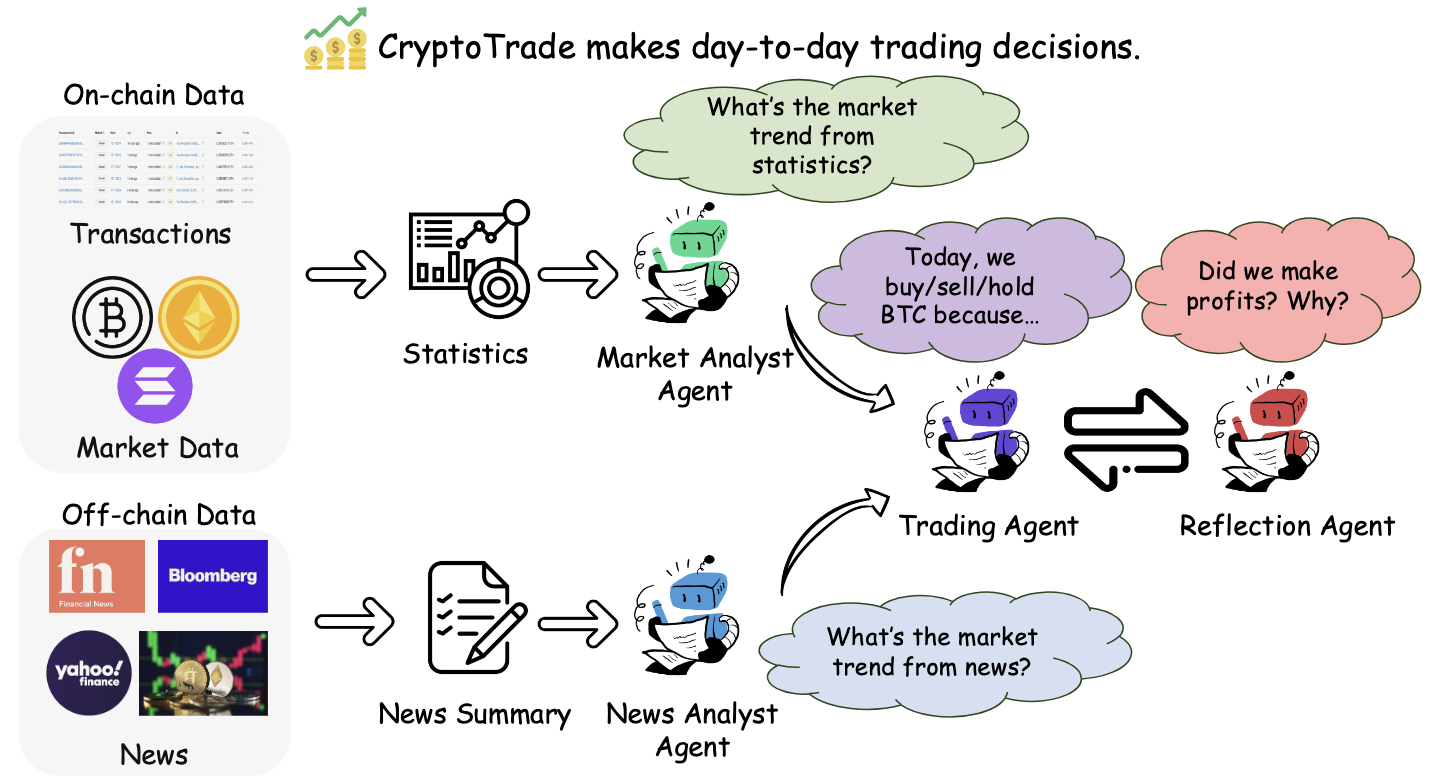}
\caption{Overview of the CryptoTrade Simulation.}
\label{fig:cryptotrade}
\end{figure}

CryptoTrade is an LLM-based trading agent designed to enhance cryptocurrency market trading by integrating both on-chain and off-chain data analysis. It leverages the transparency and immutability of on-chain data, along with the timeliness and influence of off-chain signals, such as news, to offer a comprehensive view of the market. CryptoTrade also incorporates a reflective mechanism that refines its daily trading decisions by assessing the outcomes of previous trades. \textbf{It simulates the buy and sell behaviors of human traders in the cryptocurrency market.} An overview of this simulation is shown in Figure~\ref{fig:cryptotrade} \citep{li2024cryptotrade}.

And the result of this simulation on the Ethereum market compared with other trading baselines is shown in the figure below \citep{li2024cryptotrade}.

\begin{figure}[htbp]
\centering
\includegraphics[width=0.95\textwidth]{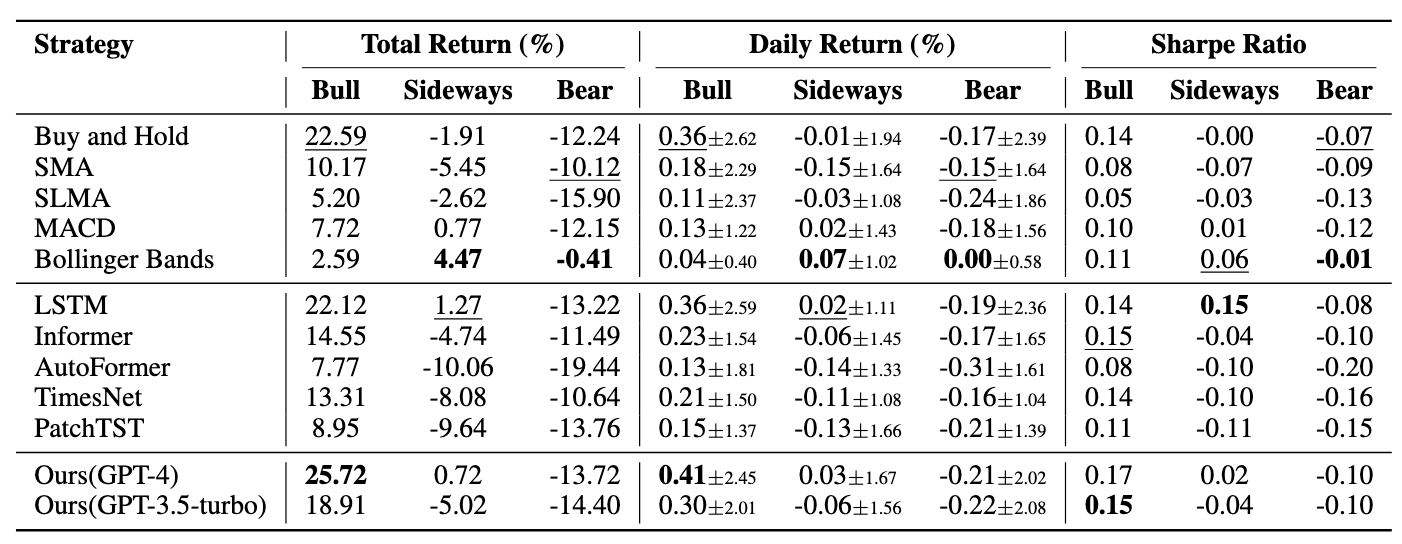}
\caption{Comparison of CryptoTrade with other trading baselines.}
\label{fig:eth_results}
\end{figure}

\begin{figure}[htbp]
\centering
\includegraphics[width=0.6\textwidth]{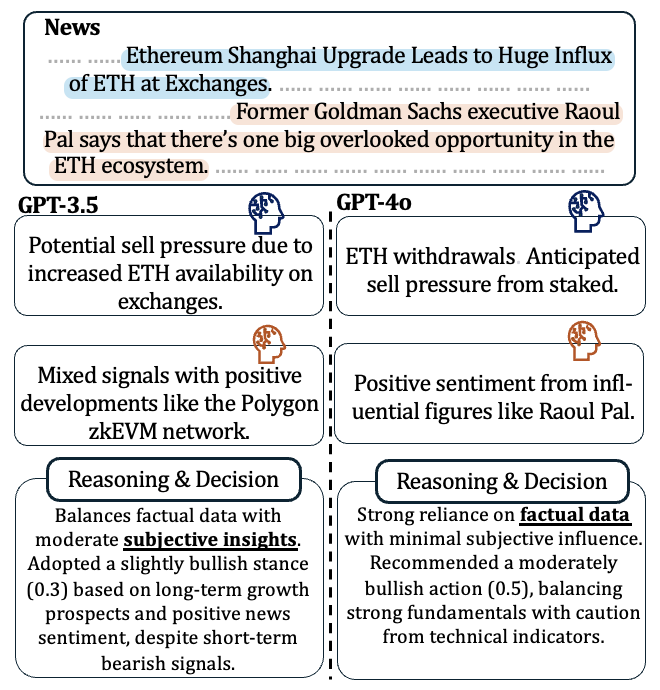}
\caption{Reasoning process of GPT-3.5 and GPT-4.}
\label{fig:reasoning}
\end{figure}

To gain deeper insights into why CryptoTrade takes specific actions, we extract the reasoning process from its simulation logs in Figure~\ref{fig:reasoning}. These logs reveal how GPT-3.5 and GPT-4o respond to the same news event: Ethereum Shanghai Upgrade.

\subsection{Key Observations}
We summarize the key observations of the CryptoTrade simulation performance and reasoning processes as follows:

\begin{enumerate}
\item \textbf{LLM Simulation Can't Outperform Buy and Hold}: In a bear market, CryptoTrade \textbf{lags behind the Buy and Hold strategy by approximately 2\%}, highlighting a significant limitation. While LLMs are expected to outperform human traders, the results do not align with this expectation.
\item \textbf{Inherent Bias}: During trading, CryptoTrade exhibited a tendency to prioritize factual information signals over sentiment-based information. While this approach can be advantageous in a bull market, it proves less effective in a bear market. For instance, in Ethereum trading, CryptoTrade outperformed the Buy and Hold strategy by 3\%, likely due to its \textbf{inherent factual bias}. However, this bias is less suited for bear markets, where profitability often requires selling assets proactively at the first signs of a downturn in the social media.
\item \textbf{Herd Behavior}: When multiple simulation agents in CryptoTrade relied on the same LLM-based models, they often made identical decisions, which could \textbf{amplify market movements rather than creating realistic market dynamics}.
\end{enumerate}

\subsection{Lessons Learned}
This case study provides several insights about LLM simulations:

\begin{enumerate}
\item \textbf{Hybrid Approaches Needed}: The most effective simulations might combine LLM agents with \textbf{some form of human oversight or intervention}, which can be injected as the format of RAG, especially for handling extreme market conditions.
\item \textbf{Bias Mitigation}: To enable LLM simulations to better replicate realistic human behaviors, it is essential to address the factual preference biases inherent in LLMs and to incorporate societal knowledge and values into their design and training.
\item \textbf{Evaluation Metrics}: Currently, the evaluation metric is solely focused on return-related mathematical metrics in trading. However, what if different individuals prefer different trading styles or strategies? How can we assess the performance of LLM simulations in such scenarios? If we aim to simulate a realistic cryptocurrency market with diverse traders, what evaluation metrics should be used?
\end{enumerate}

\section{Conclusion}
This paper highlights the limitations of LLM simulations in aligning with human behavior, encouraging deeper reflection on their ability to model the complexity of human societies. At present, LLM-based simulations offer significant potential for research, combining cost efficiency, flexibility, and the capacity to model intricate societal dynamics in innovative ways. However, addressing ethical concerns, such as bias and representation, is essential to ensure these simulations contribute \textbf{positively and equitably to our understanding of human behavior}. To better align LLM simulations with human societies, future research should focus on mitigating inherent biases, enhancing personalization, creating realistic environments, and developing reliable metrics to produce more authentic and impactful simulations.

\bibliography{iclr2025_conference}
\bibliographystyle{iclr2025_conference}

\end{document}